\documentclass[twocolumn,prb]{revtex4}
\usepackage{epsfig,amssymb,amsmath}

\usepackage{graphicx}

\usepackage{epstopdf}

\usepackage{epsfig}

\usepackage{amsmath}
\usepackage[T1]{fontenc}
\usepackage{float}
\usepackage{dcolumn}
\usepackage{hyperref}
\usepackage{amsthm}
\usepackage{amssymb}

\begin{document}
\title{Resonance phenomena
in the annular array of underdamped Josephson junctions
}
\author{I. R. Rahmonov$^{1,2}$, J. Teki\' c$^3$, P. Mali$^4$, and A. Irie$^5$, A. Plecenik$^{6}$, and Yu. M. Shukrinov$^{1,7}$}

\address{
$^1$BLTP, JINR, Dubna, Moscow Region, 141980, Russia \\
$^2$Umarov~Physical~Technical~Institute,~TAS,~Dushanbe,~734063~Tajikistan\\
$^3$"Vin\v ca" Institute of Nuclear Sciences, Laboratory for Theoretical and Condensed Matter Physics - 020, University of Belgrade, PO Box 522, 11001 Belgrade, Serbia\\
$^4$Department of Physics, Faculty of Science, University of Novi Sad, Trg Dositeja Obradovi\' ca 4, 21000 Novi Sad, Serbia\\
$^5$Department of Electrical and Electronic Systems Engineering, Utsunomiya University, 7-1-2 Yoto, Utsunomiya 321-8585, Japan\\
$^6$Department of Experimental Physics, Comenius University in Bratislava, 814 99 Bratislava, Slovakia\\
$^7$Dubna State University, Dubna,  141980, Russia
}

\date{\today}

%==================================================================================================
\begin{abstract}
Appearance and origin of resonance phenomena have been studied in the annular system of underdamped Josephson junctions.
If no fluxon is trapped in the system, dynamics is governed by the motion of fluxon-antifluxon pairs, while if trapped fluxons are present, they can move solely but also simultaneously with the pairs.
Locking between the rotating excitations (fluxons and antifluxons) and the Josephson frequency leads to the appearance of zero field steps in the current-voltage characteristics, which can
further exhibit branching due to resonance between the rotating excitations and plasma oscillations in their tale.
The number of zero field steps and their branching are strongly determined by the total number of excitations present in the system.
High resolution analysis further reveals not only some interesting properties of zero field steps,
but also shows that the current-voltage characteristics is determined not only by the number, but also by the type of excitations, i.e., whether the dynamics is governed only by the motion of fluxon-antifluxon pairs or the trapped fluxons, or they move simultaneously in the system.
\end{abstract}
%==================================================================================================

\maketitle
%==================================================================
\section{Introduction}

Resonance phenomena in Josephson junction (JJ) systems have been for years an active research topic for science and technology.
From the fundamental point of view  the JJs are an excellent devices for the studies of nonlinear dynamics of discrete systems as they represent an experimental realization of the Frenkel-Kontorova model (discrete sine-Gordon model)~\cite{Mazo, OBBook, ACFK, FlorAP, Fult, McL, Lucci15}.
At the same time from the point of their applications, JJs are promising devices for development of various fields from generation and detection of electromagnetic radiations in very low terahertz range~\cite{Mazo}, development of quantum information technologies~\cite{Clarke, WallNat, Fed, Likharev91, Herr11, Mukhanov11, Ren11, Semenov03, Volkmann13, Cirillo85, Takeuchi}, superconducting metamaterials \cite{Kis} to the fields as far as biology~\cite{Crott}.

The systems described by the sine-Gordon equations can exhibit three types of dynamical solutions: small-amplitude waves, solitons and breathers~\cite{Mazo, OBBook}.
In the JJ systems, the soliton solution has a real physical meaning since it corresponds to a quantum of flux trapped into a junction also called fluxon or Josephson vortex~\cite{Fult, McL}.
The idea that fluxon behaves as a particle-like solitary wave, which can be manipulated and controlled,
motivated creation of a new logic circuit by using Josephson fluxon as elementary bits of information~\cite{Likharev91, Herr11, Mukhanov11, Ren11, Semenov03, Volkmann13, Cirillo85, Takeuchi}.
In the creation of a new logic elements, particularly important are the long Josephson junctions described by continuous sine-Gordon equation, and the Josephson junctions parallel array by its discreet counterpart i.e. Frenkel-Kontorova model~\cite{OBBook, ACFK, FlorAP}.
The very good overview of the most prominent fluxon dynamics results in both natural and man-made systems can be found in Ref. \onlinecite{Mazo}.

Dynamics of long Josephson junctions have been subjects of numerous theoretical and experimental studies~\cite{Mazo, Eme, cirillo93}.
However, in long JJs, motion of fluxon strongly depends on the geometry and boundaries of the junctions, which
act as a mirror reflecting fluxon into antifluxon (this process can be viewed as a fluxon-antifluxon collision).
Thus, the current-voltage (I-V) characteristics of long JJs could be very complicated which makes studies of fluxon dynamics very challenging.
These problems led to the creation of annular Josephson junction as well as annular systems of Josephson junctions with a ring-shape geometry~\cite{DavPRL85}.
Annular junctions provide an undisturbed and very tunable fluxon motion, which today makes them an ideal system for the studies of fluxon dynamics~\cite{Pfeiffer08, Pfeiffer06, Ustinov93, UstPRL, Zant95,  Wall, Vernik, Watanabe, Nappi, Monaco}.

One of the interesting properties of JJs systems is the appearance of resonant steps in the current-voltage characteristics in the absence of any external radiation~\cite{Fult, Lomdhal, Kawamoto}.
Though the first experimental indication about their existence have been earlier~\cite{Scott},
the name zero filed steps (ZFS) for the observed resonant structures and the first theoretical explanation in terms of vortex motion inside the long junction were introduced by Fulton and Dyens in their pioneering work on long Josephson junction~\cite{Fult}.
The always present need for an undisturbed fluxon motion without collisions further motivated studies of resonance phenomena in annular systems~\cite {DavPRL85, Pfeiffer08, Pfeiffer06, Zant95, Watanabe},
where the appearance of ZFSs can be interpreted in terms of circulating motion of fluxons and antifluxons.

In this paper, we will examine the underdamped dynamics of an annular array of Josephson junctions (AAJJ)  particularly focusing on the origin and appearance of various resonance phenomena  in the absence external radiation.
In contrast to the previous studies of annular Josephson junctions, which were mainly focused on the zero field steps in the case of usually one trapped fluxon in a small range of currents and voltages~\cite {Pfeiffer08, Ustinov93}, here, we will examine the resonance phenomena of the AJJJ in various cases (without trapped fluxons, when dynamics is characterized by the motion of fluxon-antifluxon pairs, as well as in the case with trapped fluxons) in a wide range of currents and voltages in order to get full picture of dynamical behavior.
Our results show that dynamics of the AAJJ strongly depends not only on the number but also on the type of excitations, i.e., whether there are only trapped fluxons or the fluxon-antifluxon pairs in the system, or the trapped fluxons circulate simultaneously with fluxon-antifluxon pairs.

The paper is organized as follows.
The model is introduced
in Sec. II, while the simulation results are presented in Sec. III-VII.
Influence of the discreteness of the system on the appearance of ZFS is examined in Sec. III, while
their brunching is analyzed in Sec. IV.
The correlation between the current-voltage characteristics in the case without and in the case with trapped fluxons is presented in Sec. V.
The appearance of the effect, which we call a pulsating fluxon, on the current-voltage characteristics is shown in Sec.VI, while the influence of the type of excitations on the system dynamics is examined in Sec. VII.
Finally, Sec. VIII concludes the paper.

%==========================================================================================================
\section{Model}

We consider an annular parallel array of $N$ underdamped Josephson junctions~\cite{Pfeiffer08}.
The total length of a chain is  $L=Na$, where $a$ is a distance between the neighboring junctions.
In order to derive dynamical equations for a description of such discrete system,
we shall start from the equations for the continuous annular JJ.
In fact, the annular JJ is actually a long JJ with a periodic boundary conditions, and so, we will start from the equations for a long JJ.
The Josephson junction is considered to be long or short if its length is longer or shorter than the Josephson penetration depth $\lambda_{J}=\sqrt{\hbar S /(2e\mu_{0}DI_{c})}$, respectively.
Here $S$ is the surface area of superconducting layer, $e$ is electron charge,
$\mu_{0}$ is magnetic constant,
$D=2\lambda_{L}+d$ is effective magnetic thickness,
$\lambda_{L}$ is London penetration depth, and $d$ is the thickness of insulating layers.

According to the resistively and capacitively shunted junction (RCSJ) model~\cite{McCumber68, Stewart68}
the total current through the junction is a sum of the Josephson supercurrent, a quasiparticle (resistive normal) current, and a displacement (capacitive) current:
\begin{equation}
\label{currents}
I=I_s+I_{qp}+I_d,
\end{equation}
$$
I_{s}=I_{c}\sin\varphi,\hspace{0.2cm} I_{qp}=\frac{V}{R}\hspace{0.2cm} \text{and}\hspace{0.2cm} I_{d}=C\frac{dV}{dt},
$$
where $\varphi$ and $V$ are the phase difference and voltage across the junction, while
$R$ and $C$ are resistance and capacitance of the JJ, respectively.
The voltage $V$ is given by the Josephson relation:
\begin{equation}
\label{JR}
V=\frac{\hbar}{2e}\frac{d\varphi}{dt}=\frac{\hbar}{2e}\omega _J,
\end{equation}
where $\omega _J$ is the Josephson frequency.
In the case of long JJ, the surface current $I_{sf}$ along the superconducting layer given as
\begin{equation}
\label{surf_current}
I_{sf}=I_{c}\lambda_{J}^{2}\frac{\partial^{2}\varphi}{\partial x^{2}},
\end{equation}
should also be taken into account.

Using the Eq. (\ref{currents}), (\ref{JR}) and (\ref{surf_current}),
the total or bias current $I$ through the junction can be written as
\begin{equation}
\label{total_current_unit}
\frac{\hbar C}{2e}\frac{\partial ^{2}\varphi}{\partial t^{2}}
-I_{c}\lambda_{J}^{2}\frac{\partial ^{2}\varphi}{\partial x^{2}}
+I_{c}\sin\varphi
+\frac{\hbar}{2eR}\frac{\partial \varphi}{dt}=I.
\end{equation}
In the normalized form, the equation (\ref{total_current_unit}) for the phase difference in JJ can be simplified as
\begin{equation}
\label{total_current_norm}
\frac{\partial ^{2}\varphi}{\partial t^{2}}-\frac{\partial ^{2}\varphi}{\partial x^{2}}+\sin\varphi+\alpha\frac{\partial \varphi}{\partial t}=I,
\end{equation}
where the time is normalized with respect to the inverse plasma frequency $\omega_{p}^{-1}$, $\omega_{p}=\sqrt{2\pi I_{c}/(\Phi_{0}C)}$, the coordinate is normalized with respect to $\lambda_{J}$, and the bias current $I$ with respect to the critical current $I_{c}$.
The dissipation parameter is given as $\alpha=\sqrt{\Phi_{0}/(2\pi I_{c}R^{2}C)}$
where $L_{0}$, $C$ and $R$ are the inductance, capacitance and the differential resistance of a single cell, respectively,  and $\Phi_{0}=\frac {h}{2e}$ is the flux quantum~\cite{UstPD}.
The equation (\ref{total_current_norm}) represents the well known perturbed sine-Gordon equation.
The boundary conditions for the long JJ described by the Eq. (\ref{total_current_norm})
have the form
$\partial \varphi/\partial x|_{x=0}=\partial \varphi/\partial x|_{x=L}=H_{ext}$, which for the annular case are periodic, i.e. $\varphi(x=0)=\varphi(x=L)+2\pi M$, where $H_{ext}$ is external magnetic field and $M$ represents number of trapped fluxons inside the system.

The annular system that we are considering here can be described by the discrete version of perturbed sine-Gordon equation, which is well known as the dissipative Frenkel-Kontorova model~\cite{OBBook}:
\begin{equation}
\label{fk_equation}
\frac{d ^{2}\varphi _i}{d t^{2}}-\frac{\varphi_{i+1}+2\varphi_{i}+\varphi_{i-1}}{a^{2}}+\sin\varphi _i+\alpha\frac{d \varphi _i}{d t}=I,
\end{equation}
where $\varphi_{i}$ is the phase difference across the $i$-th junction.
The coupling between the neighboring junctions is described by the constant
$\frac {1}{a^2}$, where $a=\sqrt{2\pi L_{0}I_{c}/\Phi_{0}}$ is the discreteness parameter,
i.e., distance between two junctions normalized to the $\lambda_{J}$.

The Eq. (\ref{fk_equation}) can be linearized around the solution consisting of traveling kink (fluxon) and a small linear wave (perturbation)~\cite{Pfeiffer08}.
The dispersion law for linearized waves is given as:
\begin{equation}
\label{dispersion}
\omega_{m}=\sqrt{1+\frac{4}{a^{2}}\sin^{2}\bigg(\frac{\pi m a}{L}\bigg)}.
\end{equation}
where $m$ is an integer.

In order to calculate the I-V characteristic of the AAJJ we have used the Eq. (\ref{fk_equation}) and the Josephson relation:
\begin{equation}
\label{JR_norm}
V_i=\frac{d\varphi_i}{dt}=\omega _J,
\end{equation}
where $V_i$ is the voltage of $i$th junction normalized to $V_{0}=\hbar \omega_{p}/2e$.

Our numerical simulations were performed for the periodic boundary conditions, which in discrete case have the form:
\begin{equation}
\label{bc}
\varphi_{N+1}=\varphi_{1}+2\pi M, \hspace{0.8cm} \varphi_{0}=\varphi_{N}-2\pi M,
\end{equation}
where the spatial points $i=0$ and $i=N+1$ were assumed to be equivalent to $i=N$ and $i=1$.
We have applied the well known procedure used in Ref. \onlinecite{ShukLNCS2012} and \onlinecite{ShukJETP2012}.
The algorithm for calculating the I-V characteristic consists of several stages.
First of all, for the given value of the bias current $I$ the Eqs. (\ref{fk_equation}) and (\ref{JR_norm}) are solved numerically in the time interval $[0, T_{max}]$ using the fourth order Runge-Kutta method with the corresponding boundary conditions (\ref{bc}) and the initial conditions $\varphi_i(0)=0$, $V_i(0)=0$ at $I=0$.
As a result we obtained the time dependence of $\varphi_i(t)$ and $V_i(t)$ for the fixed value of the bias current.
Then, using the expression:
\begin{equation}
\label{v_av}
\left< V_i\right> =\frac{1}{T_{max}-T_{min}}\int\limits_{T_{min}}^{T_{max}}V_i(t)dt,
\end{equation}
where $T_{min}$ is the time necessary for system to reach the steady state, we calculated the average voltage for each junction in the system.
Further, the total average voltage $V$ is obtained using the expression $V=\sum \left<V_i\right>/N$.
In this way, for the given value of bias current $I$ the corresponding voltage $V$ was found.
Next, we change the bias current for some value $\Delta I$ and repeat the above procedure in order to obtain the next point.
By repeating this procedure for every value of the current, the I-V characteristic is produced.
We note that the solutions $\varphi_i(T_{max})$ and $V_i
(T_{max})$ obtained for the time $T_{max}$ at certain value of $I$ are used as the initial condition for the calculation of next point at the value of bias current $I+\Delta I$.

The magnetic field in the array have been calculated by the expressions:
\begin{equation}
\label{H}
\displaystyle B_{1}=\frac{\varphi_{1}-\varphi_{N}+2\pi M}{a}, \hspace{0.8cm}  \displaystyle B_{i}=\frac{\varphi_{i}-\varphi_{i-1}}{a}.
\end{equation}
During our analysis, the calculation of magnetic field time dependence in JJs was often necessary in order to understand the origin of the observed features in the I-V characteristics.

%==========================================================================================================================
\section{Zero Field steps}

In the AAJJ that we consider, in the absence of any external radiation, rotating excitations (fluxons and antifluxons) are passing repeatedly through the junctions,
which leads to resonance between the circulating excitations and Josephson frequency.
The signature of this effects are the zero field steps in the I-V characteristics of the system.
Depending on the system properties and the circulating excitations, these steps can exhibits various interesting properties.

\subsection{Zero field steps in the systems near continuum}

\begin{figure}[h!]
\centering
\includegraphics[width=80mm]{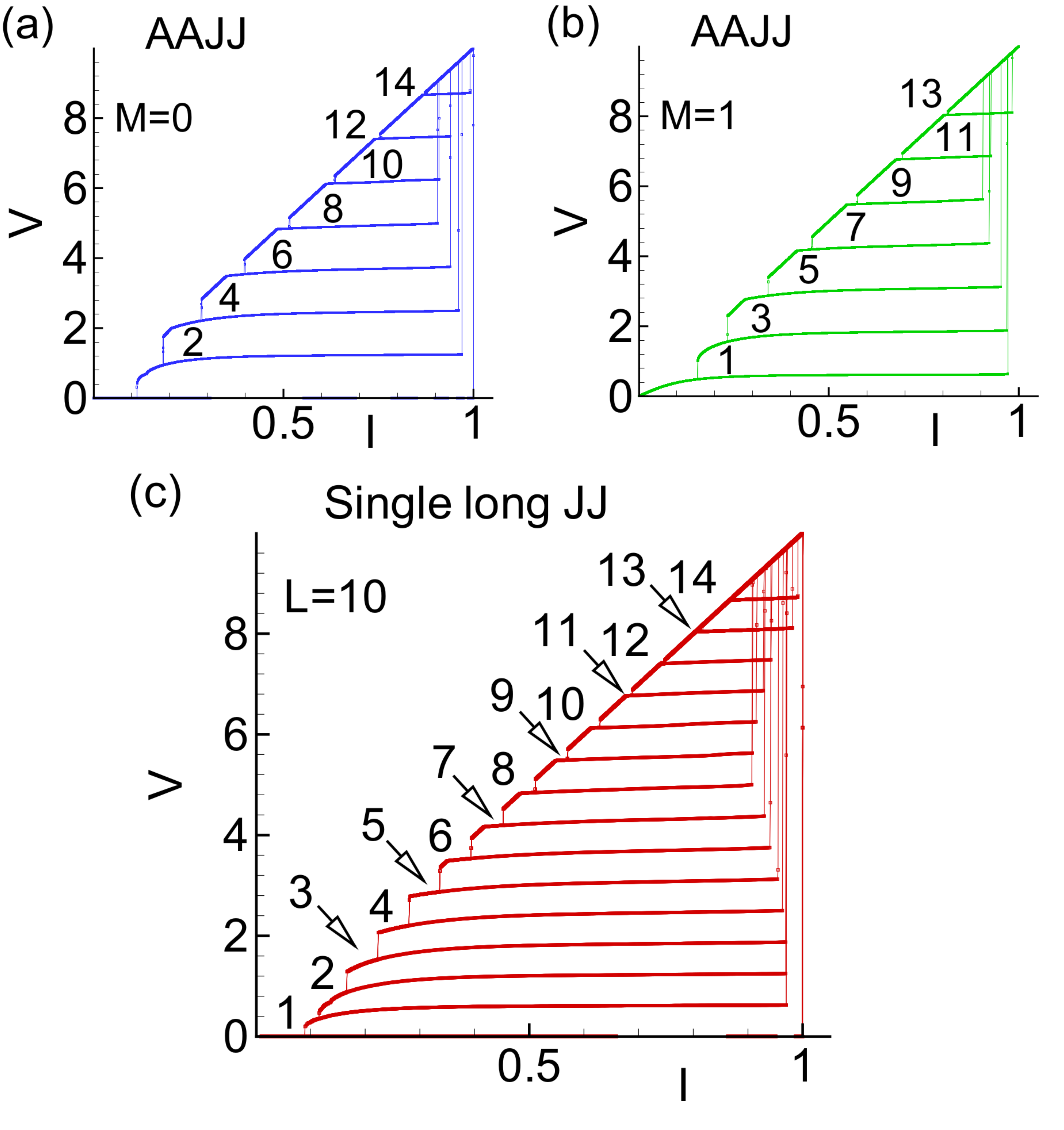}
\caption{(Color online). The current-voltage characteristics of the annular array of Josephson junctions for $N=100$, $a=0.1$, $\alpha=0.1$ and $M=0$ and $M=1$ in (a) and (b), respectively.
(c) The I-V characteristic of a single long JJ.
The numbers 2, 4, 6, and 8 mark the total number of fluxons and antifluxons in the system  $n$.}
\label{Fig1}
\end{figure}

In nonlinear systems, which exhibit resonance phenomena, discreteness plays an important role.
Let us then examine first how discreteness of the AAJJ affects the zero field steps.
We will start from the case close to the continuum limit.
If we have an AAJJ of the length $L=Na=10$, then, the near continuum limit can be achieved by placing $N=100$ junctions at the distance $a=0.1$ along the circle.
In Fig. \ref{Fig1} the current-voltage characteristic of the annular system with 100 Josephson junctions is presented for $M=0$, and $M=1$ in (a) and (b), respectively, while the result for the long Josephson junction is shown in (c) for comparison.
The presented curves are produced in the following way:
starting from zero, the current is increased until  the system transfers to the high voltage state.
Then, it is decreased to zero revealing the staircase structure,
after that it is increased again till the system is in the high voltage state.
From there, the current is decreased to the first step.
From the first step it is again increased  to the high voltage state, and
from there it is decreased to the next step.
In that way, the current was swept several times up and down, and the procedure repeated for all the steps.

These steps in Fig. \ref{Fig1} represent the well known zero field steps that appear when the voltage $V$, i.e., Josephson frequency satisfies the resonant condition $\omega_J=\frac{2\pi n u}{L}$, where $u$ is a speed of moving fluxon (antifluxon).
Here, $n$ is the total number of excitations in the system, which can be written as $n=n_f+n_{af}=2n_p+M$ where  $n_f$, $n_{af}$ and $n_p$ represent the number of fluxons, antifluxons and fluxon-antifluxons pairs (FAP), respectively.

The AAJJ is a topologically closed system, therefore the number of trapped fluxons $M$ must be conserved and new fluxons can be created only in the form of fluxon-antifluxon pairs~\cite{Monaco}.
In the case when $M=0$ (there are no trapped fluxons), dynamics is characterized by the motion of fluxon-antifluxon pairs.
Thus, in Fig. \ref{Fig1} (a)
the system exhibits only even mode resonances, where $n$ is an even number, which corresponds to the number of fluxons and antifluxons created in pairs in the system.
If we introduce one fluxon, in which case $M=1$, we can see  in Fig. \ref{Fig1} (b) that only the odd mode resonances appear, where $n$ corresponds to the sum the trapped fluxon and the number of fluxons and antifluxons, which come in pairs.
On the other hand, the long JJ in Fig. \ref{Fig1} (c) is not topologically closed, and it can exhibit both even and odd mode resonances.

%==========================================================================================================================
\subsection{Zero field steps in discrete systems}

\begin{figure}[h!]
\centering
\includegraphics[width=80mm]{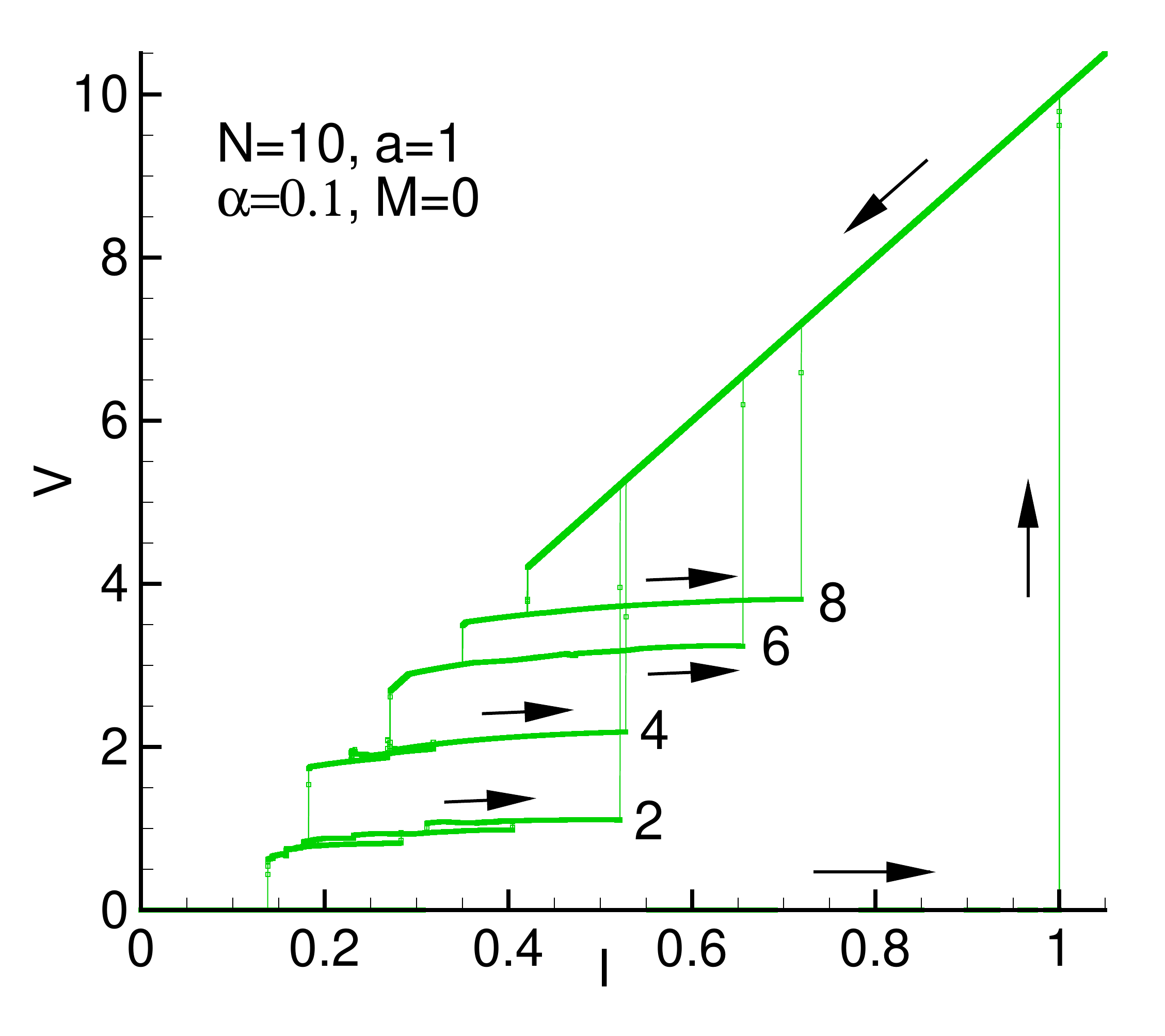}
\caption{The current-voltage characteristic of the annular array of Josephson junctions for $N=10$, $a=1$, $\alpha=0.1$ and $M=0$.
The numbers mark the total number of fluxons and antifluxons.
The arrows show direction in which the current was changed.
}
\label{Fig2}
\end{figure}

In highly discrete systems the fluxon dynamics changes significantly.
In order to discretize our system, we reduce the number of junctions to $N=10$, while keeping the same total length $L=10$ ($a=1$) as in previous case.
In Fig. \ref{Fig2} the I-V characteristics for the case $M=0$, when there are no fluxons trapped in the system, is presented.
The curves are produced following the same procedure as in Fig. \ref{Fig1}.
In this case, the resonances come from the motion of fluxon-antifluxon pairs, and the four ZFSs  correspond to $n=2, 4, 6$ and $8$.
If we compare this result with the one in  Fig. \ref{Fig1} (a), we can see that in the highly discrete case the number of steps is significantly reduced.
This reduction of resonances comes from the fact that in order for a resonance to appear there should be a certain balance between  the number of rotating excitations and the junctions in the system, i.e., the existence of stable fluxon-antifluxon pairs requires that the system is sufficiently large to permit the pair dynamics~\cite{Nappi}.
When the number of rotating fluxons and antifluxons increases, the time interval between their two consecutive passages is getting reduced.
Consequently they are constantly passing through the junctions with the frequency much higher than $\omega _J$ leaving  no time for any resonance to appear.
Further, we will present some of the interesting phenomena that we observed in our examination of zero field steps.

%==========================================================================================================================
\section{Branching of the zero field steps}

\begin{figure}[h!]
\centering
\includegraphics[width=80mm]{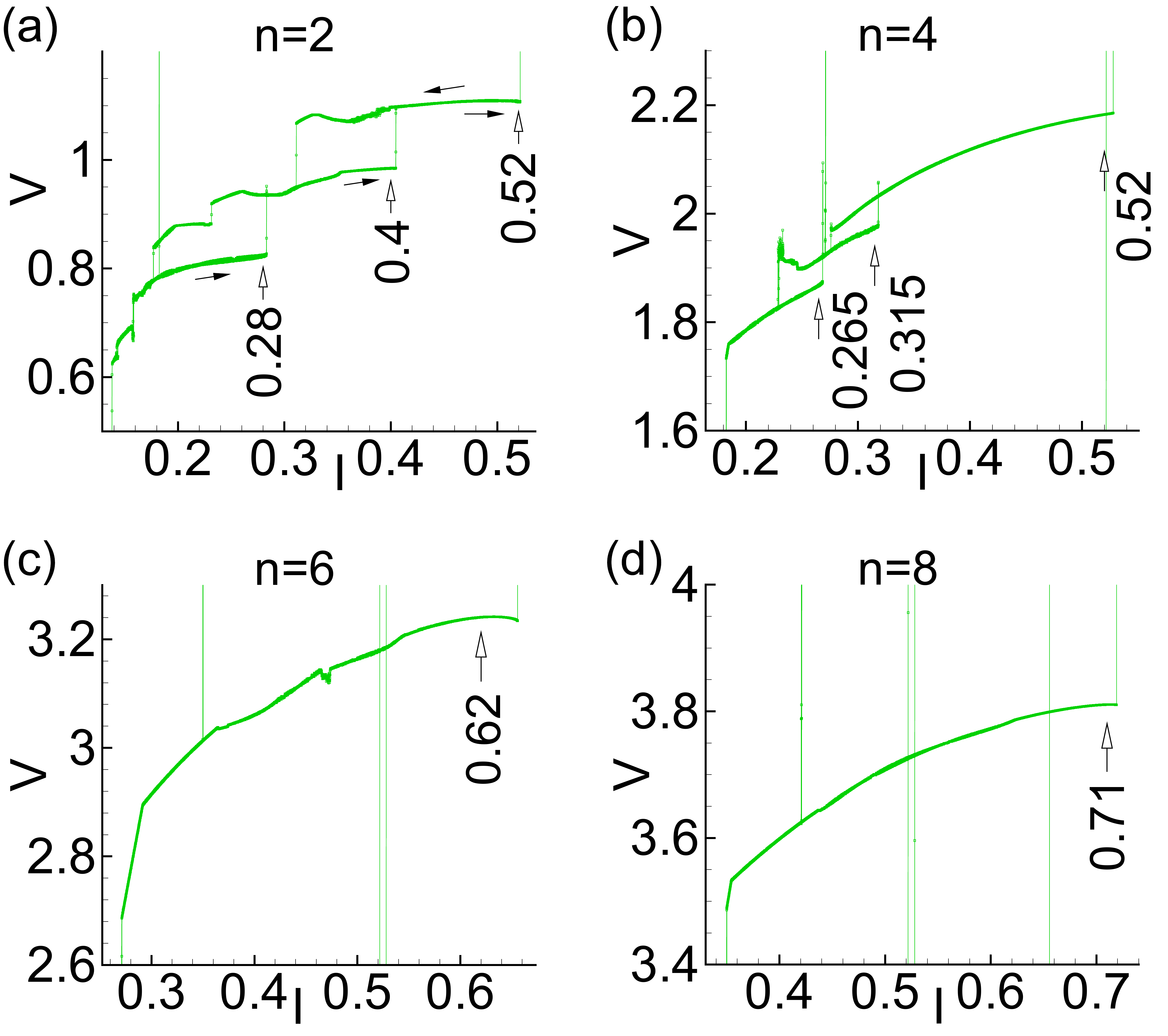}
\caption{High resolution plots
of the ZFSs shown in Fig. \ref{Fig2} for $n=2, 4, 6$ and $8$ are presented in (a), (b), (c), and (d), respectively.
The arrows mark the value of $I$ at the end of each branch.
The rest of parameters are the same as in Fig. \ref{Fig2}.
}
\label{Fig3}
\end{figure}

If we perform the high resolution analysis of the zero field steps in Fig. \ref{Fig2}, a complex structure is revealed.
In Fig. \ref{Fig3}, the enlarged regions of the I-V curves corresponding to the each of four zero field steps are presented in (a) - (d), respectively.
The first two steps in Fig. \ref{Fig3} (a) and (b) exhibit {\it branching}.
Namely, an excitations moving through the system excites small-amplitude linear waves in its tale, which are called Josephson plasma waves due to their plasma-type dispersion relation (see Eq. \ref{dispersion}).
The resonance appears due to frequency locking between the moving excitation and the small-amplitude oscillations or plasma waves (created by its motion), which results in the appearance of a series of branches at the zero field steps in the I-V characteristics of the junction.
While the ZFSs appear in both continuous and discrete systems, branching can be observed only in discrete ones~\cite{Pfeiffer08, Ustinov93}.
When $M=0$,
the dynamics is characterized by the motion of FAPs, i.e.,
the equal number of fluxons and antifluxons are circling in the opposite direction to each other along the AAJJ.
The branch of order $m$ appears, when two consecutive passing of excitations through a junction corresponds to the $m$th oscillation of the plasma (linear) wave.
As we can see in Fig. \ref{Fig3} (a), for $n=2$, where one pair, i.e., one fluxon and one antifluxon rotate, the ZFS besides the main branch, which corresponds to the frequency
$\omega_{J}=1.107$,
has also additional branches, corresponding to the frequencies $\omega_{J}=0.985, 0.882, 0.823, 0.69$, and $0.635$.
For two FAPs  in the system in Fig. \ref{Fig3} (b), the position of the ZFS $n=4$  corresponds to the frequency $\omega_{J}=2.185$, with additional two branches corresponding to the frequencies $\omega=1.977$ and $1.87$.
On the other hand, in Fig. \ref{Fig3} (c) and (d) no branching appears in the case of ZFSs with $n=6$  $(\omega_{J}=3.234)$ and $n=8$ $(\omega_{J}=3.81)$, respectively.

\begin{figure}[h!]
\centering
\includegraphics[width=80mm]{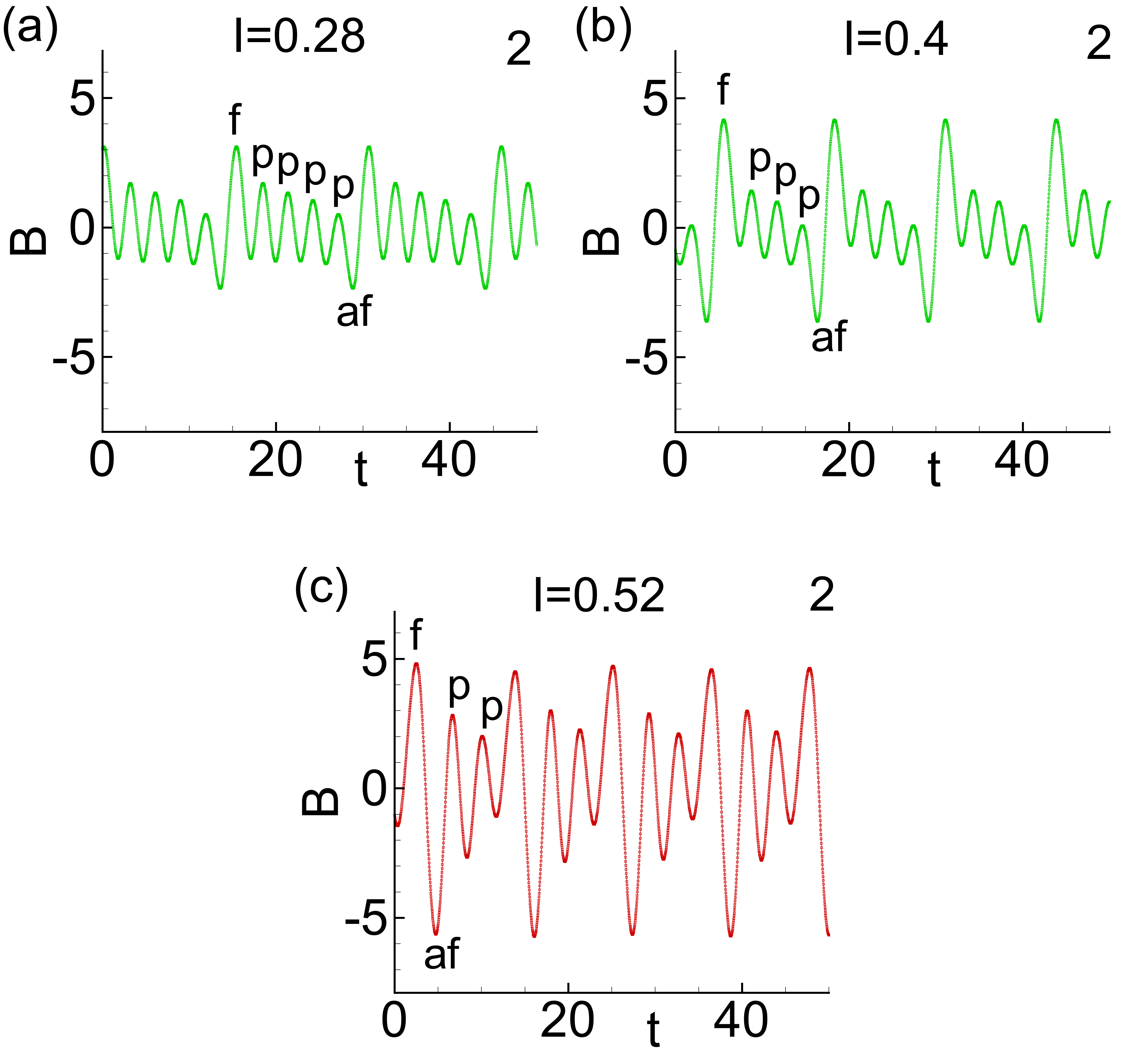}
\caption{The time dependence of the magnetic field $B$ corresponding to the first three branches of the $n=2$ ZFS with $\omega=1.107$, $0.985$, and $0.823$ measured at $I=0.28$, $0.4$ and $0.52$ in (a), (b), and (c), respectively.
The oscillations corresponding to the fluxon, antifluxon and plasma mode are marked with $f$, $af$ and $p$.
}
\label{Fig4}
\end{figure}

Let us now examine the three highest branches $\omega _J=1.107, 0.985$, and $0.823$ in Fig. \ref{Fig3} (a), and
analyze the time dependence of the magnetic field $B$ at the end of each branch for the values of the current $I$
marked by arrows.
The time dependence of magnetic field $B$ at different branches of the $n=2$ ZFS is
presented in Fig. \ref{Fig4}.
For the lowest branch in Fig. \ref{Fig4} (a),
we could see that each consecutive passage of fluxon (antifluxon) corresponds to the $5$th plasma oscillation.
At the next branch in Fig. \ref{Fig4} (b),
as the speed and the frequency of moving fluxon increases, the time between two consecutive passages of excitations  decreases, and
the branch appears due to the resonance with the $4$th plasma oscillation.
In Fig. \ref{Fig4} (c), which corresponds to the third branch,
the speed of fluxon and antifluxon increases further so that their motion locked with the third plasma oscillation.

\begin{figure}[htb]
\centering
\includegraphics[width=80mm]{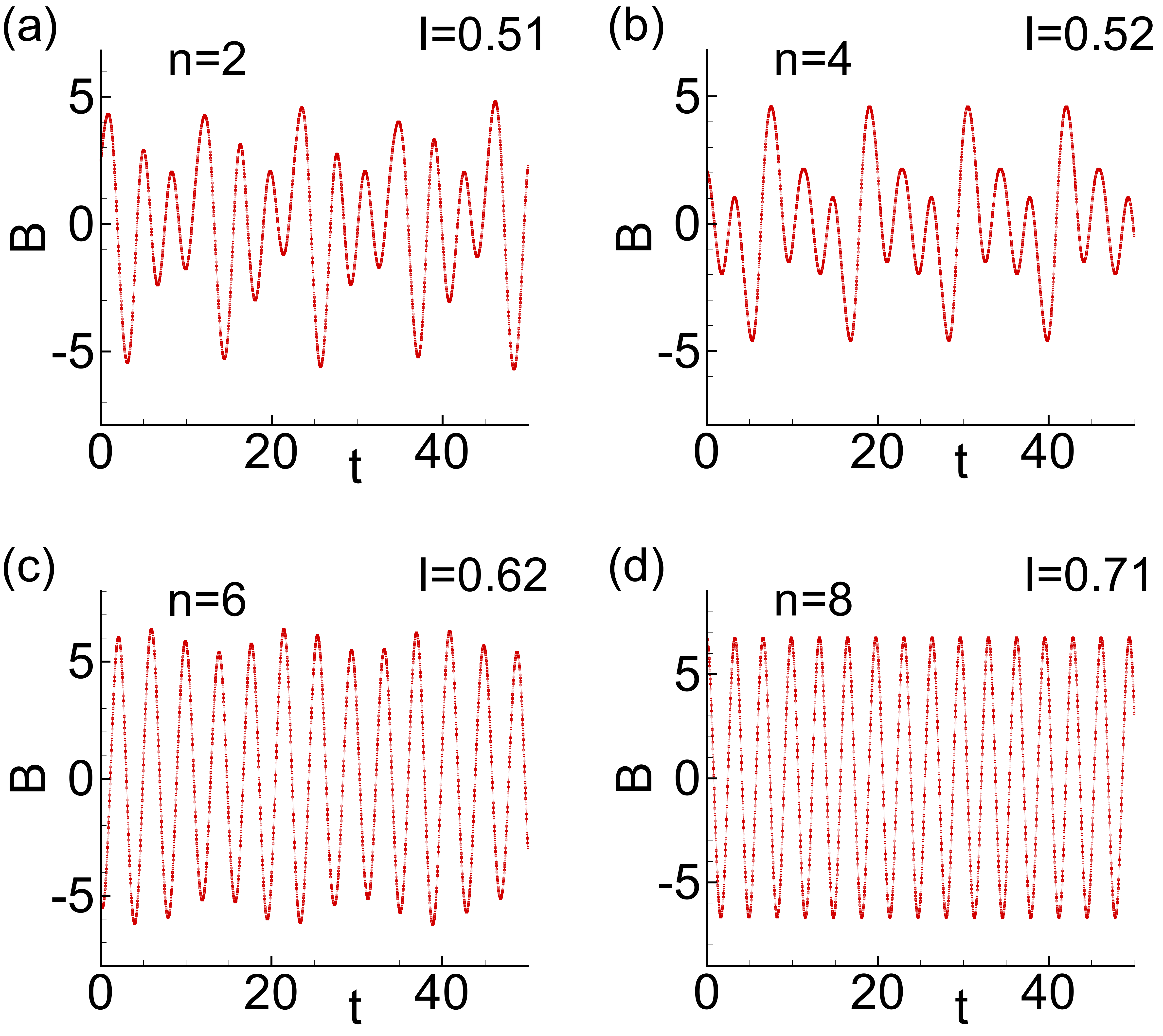}
\caption{The time dependence of the magnetic field $B$ at the zero field steps $n=2, 4, 6$, and $8$ in (a), (b), (c) and (d), respectively,
measured at the value of current $I$, which corresponds to the resonant point at the end of each step.
}
\label{Fig5}
\end{figure}

Examination of the time dependence of the magnetic field can also give us an answer, why the branching is most prominent for $n=2$ in Fig. \ref{Fig3} (a), and as the number of pairs, i.e., fluxons and antifluxons, increases to $n=4$ in Fig. \ref{Fig3} (b), the number of branches decreases, and completely disappears for $n=6$ and $8$ in Fig. \ref{Fig3} (c) and (d), respectively.
In Fig. \ref{Fig5}, the time dependence of magnetic field $B$ corresponding to zero field steps for $n=2, 4, 6,$ and $8$ is presented.
The magnetic field is measured at the resonant point (the end) of each step.
If only one or two fluxon-antifluxon pairs are moving, such as in the Fig. \ref{Fig5} (a) and (b), respectively, the time between each passage of fluxon or antifluxon is long enough, so that plasma oscillations can synchronize with their motion.
As the number of fluxon-antifluxon pairs increases, the time between each passage is getting more and more reduced.
Large number of fluxons and antifluxons will therefore leave no time for plasma oscillations between each of their passage as we can see in Fig. \ref{Fig5} (c) and (d), and consequently, there would be no resonance.
Resonances between the fluxon and plasma waves have been studied previously in annular system of underdamped Josephson junctions with one trapped fluxon~\cite{Pfeiffer08}, where the analysis was focused only on the region of the I-V characteristics, which corresponds to the first step, i.e., one circulating fluxon, and the examination of the voltage time dependence was performed.
In our case, instead of voltage it is more appropriate to use magnetic field time dependence since in that case fluxon can be distinguished from antifluxon.

%==========================================================================================================================
\section{The correlation between the steps for $M=0$ and $M=1$}

\begin{figure}[h!]
\centering
\includegraphics[width=80mm]{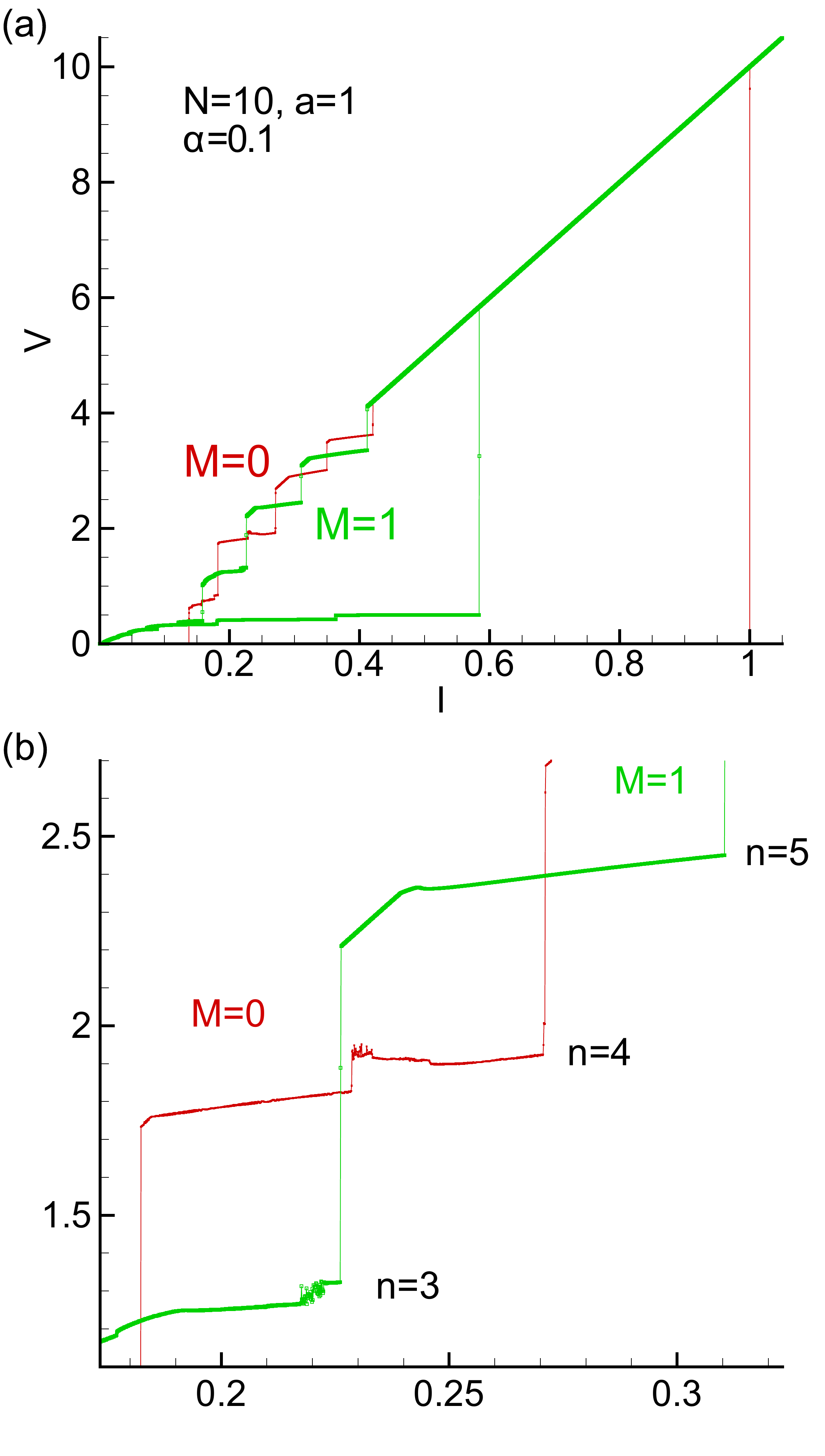}
\caption{(Color online). (a) The current-voltage characteristics of the annular array of Josephson junctions for $N=10$, $a=0.1$, $\alpha=0.1$, $M=0$ (red) and $M=1$ (green).
(b) High resolution plot of the cross section between the two curves.
}
\label{Fig6}
\end{figure}

So far, we presented only the results obtained for $M=0$, but we have also performed analysis for the systems with trapped fluxons $M\neq 0$.
In Fig. \ref{Fig6} the current-voltage characteristics for the case $M=0$ and $M=1$ are presented.
As in the case $M=0$, for $M=1$, in Fig. \ref{Fig6} (a) four zero field steps will also appear, but this time only for the odd values of $n$ ($n=1, 3, 5$, and $7$), so that $n=1$ corresponds to the one rotating fluxon, $n=3$ - to the one fluxon and one fluxon-antifluxon pair, etc.
In the high voltage state there is no difference between the curves for $M=0$ and $M=1$, however
in the region of ZFSs the curves intersect each other.
The steps for even values of $n$ $(M=0)$ alternate with the steps odd values of $n$ $(M=1)$ coming in between each other.

Interestingly, the examination of the I-V characteristics of the AAJJ with one trapped fluxon $(M=1)$ could help us understand some of the features, which appear on the ZFSs, when $M=0$.
Namely, the high resolution analysis of the I-V characteristics for $M=0$ reveals another interesting property.
In Fig. \ref{Fig6} (b), we could see some fine structure (a small peak), that appears on the step $n=4$, at $I=0.23$.
The origin of this effect could be understood from the comparative analysis of the I-V curves for both values of $M$.
As we can see,
the appearance of this defect on $n=4$ step is at the value of current ($I=0.23$), for which the I-V curve for $M=1$ changes from $n=5$ to $n=3$ state.
For given system parameters the ability of the annular Josephson junction to exhibit certain number of FAPs is determined by the current $I$.
Thus, as the current is decreased from the high voltage state, at some value of $I$, the AAJJ transfers to the state $n=8$ (4 fluxons and 4 antifluxons), and as we decrease $I$ further, we can see that at certain values of the current, the AAJJ can be in the state with $n=7$, $6$,..., $1$ excitations.
In Fig. \ref{Fig6} (b), at the step $n=4$ ($M=0$), we have 2 fluxon-antifluxon pairs moving along the AAJJ, and when the current decreases to the value $I=0.23$, the AAJJ is in the I-V region which corresponds to $n=3$ state.
However, it is impossible for AAJJ to transfer to that state due to conservation of $M$ ($n_f=n_{af}$ for $M=0$).
On the contrary, the two pairs will, due to their inertia, continue moving till the current decreases to the value, for which the system is reduced to $n=2$ state with only one fluxon-antifluxon pair.
Also in Fig. \ref{Fig6} (b), we could see that the curve for $M=1$ is slightly shifted to the right comparing to the curve $M=0$.
This might be due to inertial effect, since for $M=1$ the system contains one more fluxon in addition to fluxon-antifluxon pairs, and more fluxons will simply have more inertia.
Thus, when the current is decreased, 3 fluxons and 2 antifluxons $(n=5)$ would keep moving longer than  2 fluxons and 2 antifluxons $(n=4)$.

%==================================================================================================================
\section{Pulsating fluxon}

\begin{figure}[h!]
\centering
\includegraphics[width=80mm]{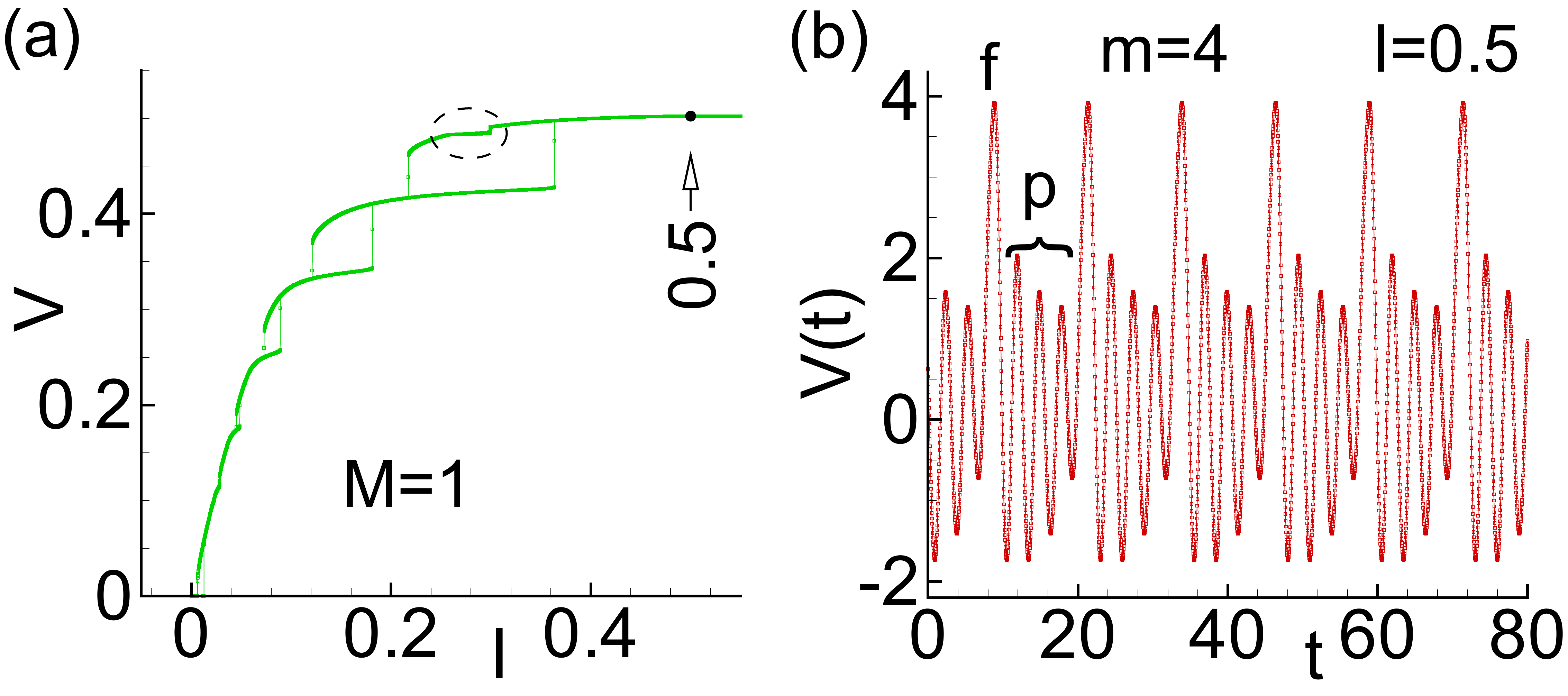}
\caption{(Color online). (a) The I-V curve of the AAJJ for $M=1$ in the region of the step $n=1$ for the one fluxon state.
(b) Voltage time dependence at the upper branch of the one fluxon zero field step for the value of current marked by arrow.
}
\label{Fig7}
\end{figure}

Let us now examine the step $n=1$ for the AAJJ with one trapped fluxon $M=1$.
The high resolution plot of the I-V curve from the Fig. \ref{Fig6} corresponding to the step $n=1$ (1 fluxon state) is presented in Fig. \ref{Fig7}.
As in the previous case  for $M=0$ in Sec. IV, now for $M=1$ in Fig. \ref{Fig7} (a) the step $n=1$ also exhibits branching.
If we analyze the voltage-time dependence, for example for a given value of $I$ marked by arrow  at the step $n=1$ in Fig. \ref{Fig7} (a), we could see in Fig. \ref{Fig7} (b), that the upper branch corresponds to the fluxon passing at the every fourth plasma oscillation.

\begin{figure}[h!]
\centering
\includegraphics[width=80mm]{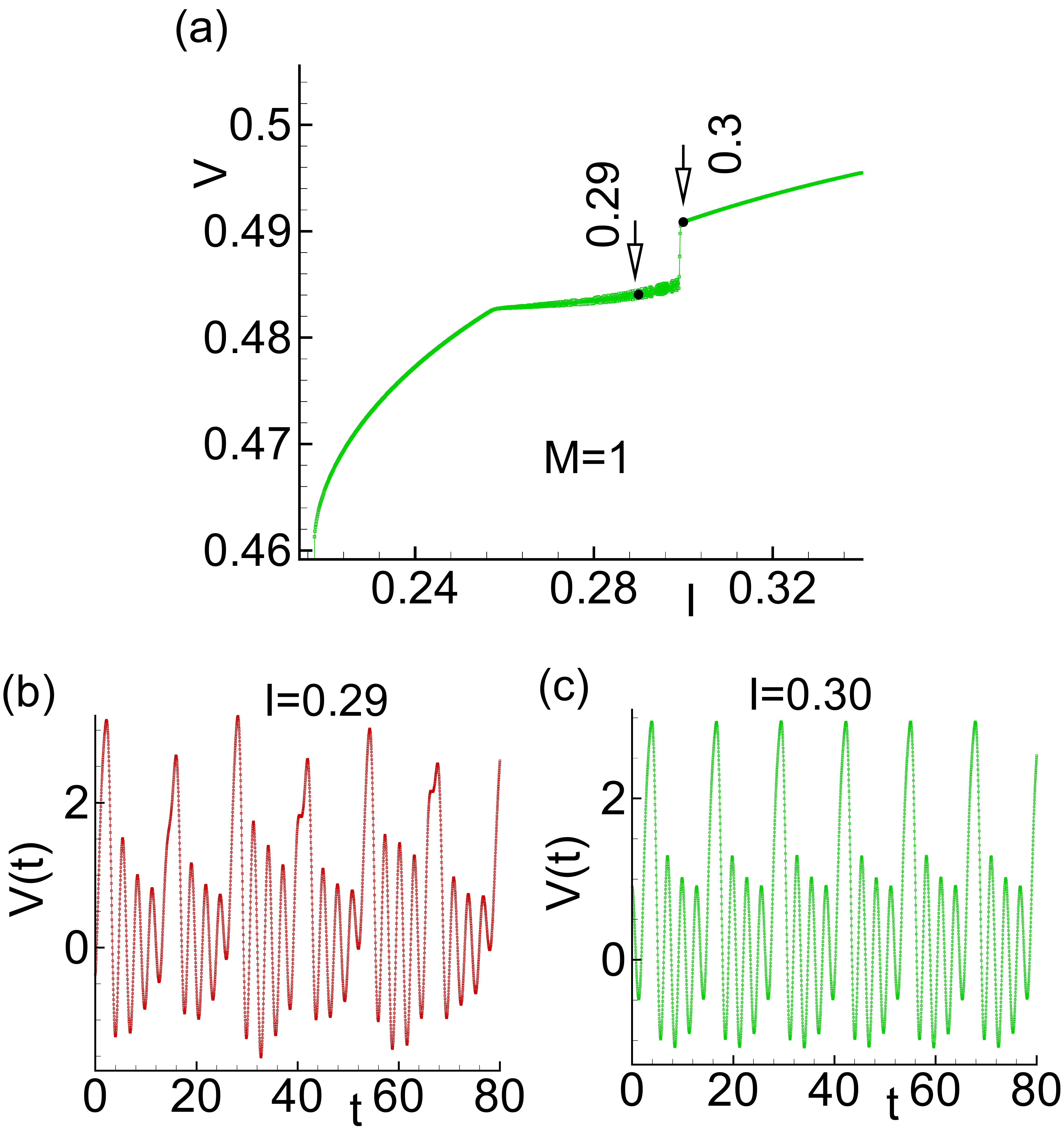}
\caption{(Color online). (a) High resolution plot of the area circled by dashed line in Fig. \ref{Fig7} (a).
Voltage time dependence at $I=0.29$ and  $0.3$ in (b) and (c) respectively.
}
\label{Fig8}
\end{figure}

However, besides branching, when $M=1$ the step in Fig. \ref{Fig7} (a) exhibits an interesting feature marked by dashed circle, which does not appear in the I-V characteristics of the AAJJ with no trapped fluxons ($M=0$).
In Fig. \ref{Fig8}, the high resolution plot of the area circled by dashed line in Fig. \ref{Fig7} (a), and the corresponding voltage time dependence at two different values of current $I$ for the AAJJ with one trapped fluxon $(n=1)$ are presented.
If we look at Fig. \ref{Fig8} (a), at first it might seem that there is an additional branch, nevertheless, the voltage time dependence at the two values of currents marked by arrows in  Fig. \ref{Fig8} (b) and (c), shows that in both cases we have resonance of order $m=4$ (fluxon passes every $4$th plasma oscillation).
However, in  Fig. \ref{Fig8} (b) for the value of current $I=0.29$, we could clearly see changes of the maxima corresponding to the fluxon, which we therefore call a {\it pulsating fluxon}.
We performed simulations for different system parameters and discovered that this phenomenon existed only in the underdamped case and disappeared as damping was increased.

%==================================================================================================================
\section{Comparative analysis of the I-V characteristics for $M=0$ and $M=2$}

\begin{figure}[h!]
\centering
\includegraphics[width=80mm]{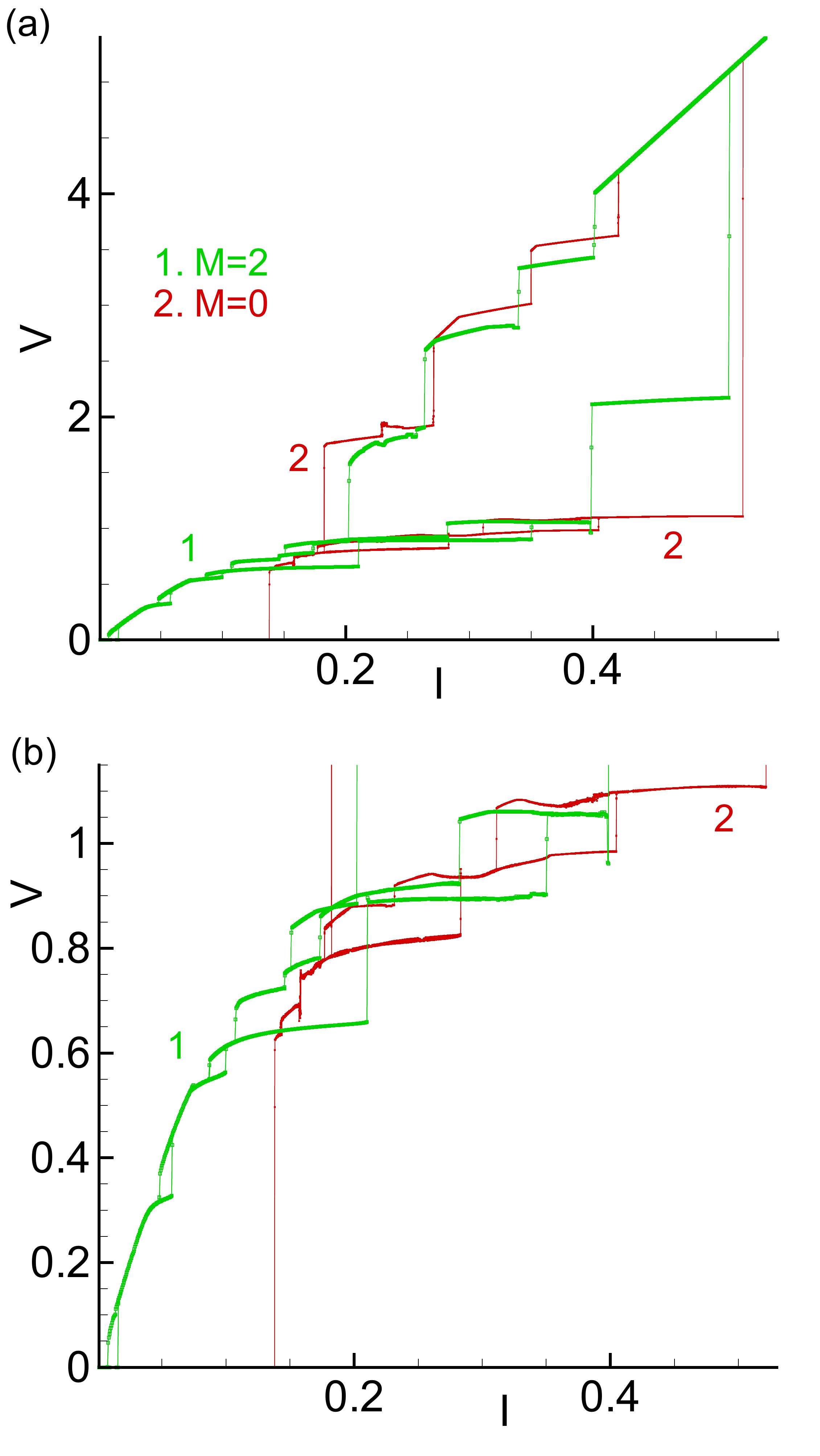}
\caption{(Color online). (a) The I-V characteristics of the AAJJ for $N=10$, $a=1$, $\alpha=0.1$, $M=0$ in red and $M=2$ in green.
(b) The high resolution plot of the $n=2$ step.
}

\label{Fig9}
\end{figure}

In order to further examine how the type of excitations affects the AAJJ, let us compare two cases:
$M=0$ and $M=2$.
At first one might expect that the I-V characteristics for $M=0$ and $M=2$ should be the same since in both cases ZFSs appear for $n=2, 4, 6,$ and $8$, however, this is not the case.
In Fig. \ref{Fig9}, the I-V characteristics for the case $M=0$ with no trapped fluxon, and the case $M=2$ with two trapped fluxons is presented.
As we can see in Fig. \ref{Fig9} (a), the steps are shifted respect to each other.
On the ZFS $n=2$, for $M=0$ we have one fluxon and one antifluxon, while for $M=2$ we have two fluxons.
On the next step $n=4$, we will have 2 fluxons and 2 antifluxons for $M=0$, while for  $M=2$ there will be 3 fluxons and one antifluxon, etc.
So, though the total number of excitations $n$ is the same for both values of $M$,
the numbers of fluxons and antifluxons are different.
For $M=0$ we always have the same number of fluxons and antifluxons $n_f=n_{af}$, while
for $M=2$ in addition to fluxon-antifluxon pairs we have 2 trapped fluxons, so that
$n_f\neq n_{af}$
and this completely change the dynamics of the system.

In the high resolution plot of the step $n=2$ presented in Fig. \ref{Fig9} (b), we can also see that for $M=2$ voltage goes to zero as $I$ goes to zero, while for $M=0$ voltage goes to zero around $I=0.14$.
This comes from that fact that for $M=0$,
fluxon-antifluxon pair can exist only above some critical current.
So, when $I$ becomes smaller than that value, there are no fluxon-antifluxon pairs, and consequently, $V$ drops to zero.
On the other hand, if we have two trapped fluxons, they will be moving as long as $I\neq 0$, and as $I$ goes to zero, their speed and consequently the voltage will go to zero.

%==========================================================================================================
\section{Conclusion}

In this work, detail analysis of the resonance phenomena in the annular array of Josephson junctions have been presented.
The observed effects and their physical origins can be summarized as follows:
in the absence of external radiation the system exhibits
zero field steps due to locking between the rotating excitations (fluxons and antifluxons) and the Josephson frequency;
these steps can further exhibit branching due to locking between the rotating excitations and plasma oscillations in their tale.
The ability of system to exhibit ZFSs and their branching is determined by the number of excitations in the system.
The branching appears always on the lower steps, and as the number of fluxons and antifluxons increases, gradually disappears at the higher steps.
Further examinations reveal not only some interesting properties of those steps, such as pulsating fluxon and correlation between the current-voltage characteristics for the cases without and with trapped fluxons, but also show
that the dynamics of AAJJ is determined not only by the number, but also by the type of excitations, i.e., the current-voltage characteristics will be completely different depending on whether there are only fluxon-antifluxon pairs or the trapped are also present in the system.
In other words, the system with one fluxon and one antifluxon, for example, does not behave the same as the system with two trapped fluxons, though the total number of excitations is the same.

Annular Josephson junctions posses an enormous potential for various technological applications~\cite{Mazo}.
Superconducting digital technology  is capable of achieving much higher energy efficiency than other technologies~\cite{Fed, Likharev91, Herr11, Mukhanov11, Ren11, Semenov03, Volkmann13, Cirillo85, Takeuchi},
and fluxon dynamics as well as resonance phenomena are in the core of some
of the most advanced ideas in those fields.
Another interesting application of annular Josephson junctions is in superconducting metamaterials with a number of unique properties, which are difficult to achieve in any other way~\cite{Kis}.
A generic element of such a material is a superconducting ring split by a Josephson junction, and one of the most recent theoretical and experimental studies have been dedicated to the resonant response of such metamaterials to the external signal in strongly nonlinear regimes~\cite{Kis}.
Regardless of the field in which the annular Josephson junctions have application, a good understanding of their dynamics is crucial.
For that reason, the resonance phenomena that we presented require further experimental examination that we will present in the future.

%============================================================================================================================================

\begin{acknowledgments}

J. Teki\' c whish to thank to Prof. Yu. M. Shukrinov and the BLTP, JINR, Dubna in Russia for their generous hospitality where a part of this work was done.
This work was supported by the Foundation for the Advancement of Theoretical Physics and Mathematics "Basis".
The reported study was funded by the RSF
research projects  18-02-00318, 18-52-45011-IND. Numerical simulation was funded by  RSF project  18-71-10095.
This work was supported by the Serbian Ministry of
Education and Science under Contracts No. OI-171009 and No. III-45010
and  by the Provincial Secretariat for High Education and Scientific Research of Vojvodina (Project No. APV 114-451-2201).
\end{acknowledgments}

%==================================================================================================================

%===============================================================================================================

\end{document}